# On Arthur Eddington's Theory of Everything

Helge Kragh[*]

**Abstract:** From 1929 to his death in 1944, A. Eddington worked on developing a highly ambitious theory of fundamental physics that covered everything in the physical world, from the tiny electron to the universe at large. His unfinished theory included abstract mathematics and spiritual philosophy in a mix which was peculiar to Eddington but hardly intelligible to other scientists. The constants of nature, which he claimed to be able to deduce purely theoretically, were of particular significance to his project. Although highly original, Eddington's attempt to provide physics with a new foundation had to some extent parallels in the ideas of other British physicists, including P. Dirac and E. A. Milne. Eddington's project was however a grand failure in so far that it was rejected by the large majority of physicists. A major reason was his unorthodox view of quantum mechanics.

## 1. Introduction

Arthur Stanley Eddington is recognized as one of the most important scientists of the first half of the twentieth century (Douglas 1956). He owes this elevated position primarily to his pioneering work in astronomy and astrophysics, and secondarily to his expositions of and contributions to the general theory of relativity. Eddington developed a standard model of the interior structure of stars and was the first to suggest nuclear reactions as the basic source of stellar energy. In 1919 he rose to public fame when he, together with Frank Dyson, confirmed Einstein's prediction of the bending of starlight around the Sun. Six years later he applied general relativity to white dwarf stars, and in 1930 he developed one of the first relativistic models of the expanding universe known as the Lemaître-Eddington model.

---

[*] Niels Bohr Institute, University of Copenhagen, 2100 Copenhagen, Denmark. E-mail: helge.kragh@nbi.ku.dk. 

Not satisfied with his accomplishments within the astronomical sciences, during the last part of his life Eddington concentrated on developing an ambitious theory of fundamental physics that unified quantum mechanics and cosmology. The present essay is concerned solely with this grand theory, which neither at the time nor later won acceptance. Although Eddington's fundamental theory has never been fully described in its historical contexts, there are several studies of it from a philosophical or a scientific point of view (e.g. Kilmister 1994; Durham 2006; French 2003). There have also been several attempts to revive scientific interest in Eddington's theory (such as Wesson 2000), but none of them have been even moderately successful. At any rate, this essay is limited to the theory's historical course and its reception among physicists.

## 2. Warming up

Although Eddington only embarked on his ambitious and lonely research programme of unifying the atom and the universe in 1929, some of the features of this programme or theory can be found in his earlier work. When Dirac's quantum equation of the electron opened his eyes to a whole new foundation of physics, he was well prepared.

The significance of combinations of the constants of nature, a key feature in the new theory, entered Eddington's masterpiece of 1923, *The Mathematical Theory of Relativity*. Referring to the "very large pure number [given] by the ratio of the radius of the electron to its gravitational mass = $3 \times 10^{42}$," Eddington (1923, p. 167) suggested a connection to "the number of particles in the world – a number presumably decided by pure accident." And in a footnote: "The square of $3 \times 10^{42}$ might well be of the same order as the total number of positive and negative electrons" (see also Eddington 1920a, p. 178). Of course, Eddington's reference to the positive electron was not to the positron but to the much heavier proton, a name introduced by Ernest Rutherford in 1920 but not generally used in the early 1920s. While Eddington in 1923 thought that the number of particles in the world was accidental, in his later theory this "cosmical number" appeared as a quantity strictly determined by theory.

A somewhat similar suggestion had been ventured by Hermann Weyl a few years earlier when considering the dimensionless ratio of the electromagnetic radius of the electron ($r_e = e^2/mc^2$) to its gravitational radius ($r_g = Gm/c^2$). According to Weyl (1919), the ratio of the radius of the universe to the electron radius was of the same order as $r_e/r_g \cong 10^{40}$, namely

$$\frac{r_e}{r_g} = \frac{e^2}{Gm^2} \approx \frac{R_E}{r_e},$$

where $R_E$ denotes the radius of the closed Einstein universe. The huge value of the electrical force relative to the gravitational force as given by $e^2/Gm^2$ had been pointed out much earlier (e.g. Davis 1904), but it was only with Weyl and Eddington that the number ca. $10^{40}$ was connected to cosmological quantities.

Another feature of Eddington's mature philosophy of physics, apart from its emphasis on the constants of nature, was the fundamental significance he ascribed to mind and consciousness. Physics, he argued, can never reveal the true nature of things but only deals with relations between observables that are subjectively selected by the human mind. As Eddington (1920b, p. 155) stated in his earliest philosophical essay, he was "inclined to attribute the whole responsibility for the laws of mechanics and gravitation to the mind, and deny the external world any share in them." By contrast, "the laws which we have hitherto been unable to fit into a rational scheme are the true natural laws inherent in the external world, and mind has no chance of moulding them in accordance with its own outlook." The same theme appeared prominently in his Gifford Lectures delivered in early 1927, where Eddington (1928, p. 281) provocatively concluded that "the substratum of everything is of mental character."

While up to this time Eddington had exclusively relied on the admired theory of general relativity, he now also appealed to the new symbolic quantum theory as a further argument in favour of his view that physicists manufacture the phenomena and laws of nature. Yet, although he took notice of Heisenberg's and Schrödinger's quantum mechanics, for a while he was uncertain of how to make sense of it within the framework of a unified theory of relativistic physics. In *The Mathematical Theory of Relativity* (1923, p. 237) he wrote: "We offer no explanation of the occurrence of electrons or of quanta. … The excluded domain forms a large part of physics, but it is one in which all explanation has apparently been baffled hitherto." Four years later the situation had not changed materially.

## 3. Eddington meets the Dirac equation

When Eddington studied Paul Dirac's relativistically invariant wave equation of the electron in early 1928 he was fascinated but also perplexed because it was not written in the language of tensor calculus (Eddington 1936, p. 2). Although he recognized the new quantum equation as a great step forward in unifying physics, he also thought that it did not go far enough and consequently decided to generalize it. For an electron (mass $m$, charge $e$) moving in a Coulomb field the positive-energy Dirac equation can be written

$$\frac{ih}{2\pi}\frac{\partial \psi}{\partial t} = \frac{e^2}{r}\psi + c\sqrt{(ih/2\pi)^2 \Delta + m^2 c^2}\,\psi,$$

where $\Delta = (\partial^2/\partial x^2, \partial^2/\partial y^2, \partial^2/\partial z^2)$ is the Laplace operator. Eddington (1929; 1931) rewrote the equation by introducing two constants, namely

$$\alpha^{-1} = \frac{hc}{2\pi e^2} \quad \text{and} \quad \gamma = \frac{2mc}{h},$$

In this way he arrived at

$$-\alpha \frac{\partial \psi}{\partial t} = \frac{ic}{r}\psi + c\sqrt{\alpha^2 \Delta - \gamma^2}\,\psi$$

The constant $\alpha^{-1}$ is the inverse of what is normally called the fine-structure constant, but Eddington always (and somewhat confusingly) reserved the name and symbol $\alpha$ for the quantity $hc/2\pi e^2$. In his initial paper of 1929 he applied his version of the Dirac equation and his unorthodox understanding of Pauli's exclusion principle to derive the value

$$\alpha^{-1} = 16 + \tfrac{1}{2} \times 16 \times (16 - 1) = 136$$

By 1929 the fine-structure constant was far from new, but it was only with Eddington's work that the dimensionless combination of constants of nature was elevated from an empirical quantity appearing in spectroscopy to a truly fundamental constant (Kragh 2003). Moreover, Eddington was the first to focus on its inverse value and to suggest – indeed to insist – that it must be a whole number. He was also the first to argue that $\alpha$ was of deep cosmological significance and that it should be derivable from fundamental theory.

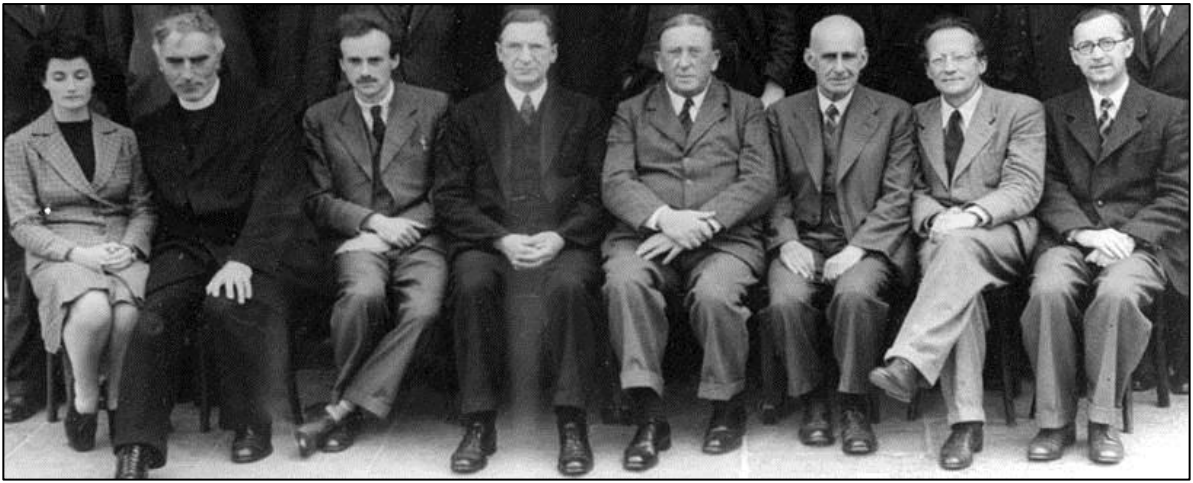

**Fig. 1.** Dirac (third from left) with Eddington and Schrödinger (third and second from right) at a colloquium in 1942 at the Dublin Institute for Advanced Studies. The other persons, from the left, are Sheila Power, Pádraig de Brún, Eamon de Valera, Arthur Conway, and Albert J. McConnell. The Dublin Institute was established in 1940 by de Valera, the Irish political leader who was trained in mathematics and much interested in the sciences. Source: http://www-history.mcs.st-and.ac.uk/Biographies/Tinney.html.

When Eddington realized that the theoretical value $\alpha^{-1} = 136$ did not agree with experiment, at first he pretended to be undisturbed. "I cannot persuade myself that the fault lies with the theory," he wrote in his paper of 1929. All the same, as experiments consistently showed that $\alpha^{-1} \cong 137$ he was forced to look for a fault in his original theory. He soon came to the conclusion that $\alpha^{-1} = 137$ *exactly*, arguing that the extra unit was a consequence of the exclusion principle, which in his interpretation implied the indistinguishability of any pair of elementary particles in the universe. For the rest of his life Eddington (1932, p. 41) stuck to the value 137, which he claimed to have "obtained by pure deduction, employing only hypotheses already accepted as fundamental in wave mechanics."

It should be pointed out that although Dirac's linear wave equation served as the inspiration for Eddington's grand theory, for him it was merely a temporary stepping stone towards a higher goal. He felt that relativistic quantum mechanics, whether in Dirac's version or some of the other established versions, was still characterized by semi-empirical methods that prevented a truly rational foundation of laws unifying the micro-cosmos and the macro-cosmos. In his monograph *Relativity Theory of Protons and Electrons* (1936, pp. 6-7) Eddington emphasized the difference between his theory and the one of Dirac:

> Although the present theory owes much to Dirac's theory of the electron, to the general coordination of quantum theory achieved in his book *Quantum Mechanics* … it is not "Dirac's theory"; and indeed it differs fundamentally [from it] on most points which concern relativity. It is definitely opposed to what has commonly been called "relativistic quantum theory," which, I think, is largely based on a false conception of the principles of relativity theory.

While Dirac and other specialists in quantum mechanics disagreed with the last part of the equation, they very much agreed with the first part.

## 4. Constants of nature

The fine-structure constant was not the only constant of nature that attracted the attention of Eddington. On the contrary, he was obsessed by the fundamental constants of nature, which he conceived as the building blocks of the universe and compared to the notes making up a musical scale: "We may … look on the universe as a symphony played on seven primitive constants as music played on the seven notes of a scale" (Eddington 1935, p. 227). The recognition of the importance of constants of nature is of relatively recent origin, going back to the 1880s, and Eddington was instrumental in raising them to the significance they have in modern physics (Barrow 2002; Kragh 2011, pp. 93-99). Whereas the *fundamental* constants, such as the mass of the electron and Planck's quantum constant are generally conceived to be irreducible and essentially contingent quantities, according to Eddington this was not the case. Not only did he believe that their numerical values could be calculated, he also believed that he had succeeded in actually calculating them from a purely theoretical basis (see Fig. 2). For example, for Newton's gravitational constant he deduced $G = 6.6665 \times 10^{-11}$ N m$^2$ kg$^{-2}$, in excellent agreement with the experimental value known at the time, which he stated as $(6.670 \pm 0.005) \times 10^{-11}$ N m$^2$ kg$^{-2}$ (Eddington 1946, p. 105).

The constants of nature highlighted by Eddington were the mass $m$ and charge $e$ of the electron, the mass of the proton $M$, Planck's constant $h$, the speed of light $c$, the gravitational constant $G$, and the cosmological constant $\Lambda$. To these he added the cosmical number $N^*$ and, on some occasions, the number of dimensions of space-time (3 + 1). Two of the constants were original to him and deserve mention for this reason. The cosmological constant introduced by Einstein in his field equations of 1917 was not normally considered a constant of nature and was in

| Constant | Calculated | Observed |
|---|---|---|
| Mass of electron | $9.10924 \times 10^{-28}$ g | 9.1066 |
| Mass of proton | $1.67277 \times 10^{-24}$ g | 1.67248 |
| Proton–electron mass ratio | 1836.34 | 1836.27 |
| Elementary charge | $4.80333 \times 10^{-10}$ esu | 4.8025 |
| Planck's constant | $6.62504 \times 10^{-27}$ erg s | 6.6242 |
| Inverse fine-structure constant | 137 | 137.009 |
| Gravitational constant | $6.6665 \times 10^{-8}$ cm$^3$ g$^{-1}$ s$^2$ | $6.670 \pm 0.005$ |
| Hubble constant | 572.36 km s$^{-1}$ Mpc$^{-1}$ | 560 |
| Magnetic moment of H atom | 2.7899 nuclear magnetons | $2.7896 \pm 0.0008$ |
| Magnetic moment of neutron | 1.9371 nuclear magnetons | $1.935 \pm 0.02$ |

**Fig. 2.** Eddington's calculations of constants of nature compared to those observed in the 1940s. Reproduced from Kragh (2011), p. 99.

any case ill-regarded by many cosmologists and astronomers in the 1930s. With the recognition of the expanding universe most specialists followed Einstein in declaring the cosmological constant a mistake, meaning that $\Lambda = 0$. Eddington emphatically disagreed. He was convinced that the $\Lambda$-constant was indispensable and of fundamental importance, for other reasons because he conceived it as a measure of the repulsive force causing the expansion of the universe. Appealing to Einstein's original relation between the constant and the radius of the static universe,

$$\Lambda = \frac{1}{R_\mathrm{E}^2} ,$$

he considered $\Lambda$ to be the cosmic yardstick fixing a radius for spherical space. "I would as soon think of reverting to Newtonian theory as of dropping the cosmical constant," Eddington (1933, p. 24) wrote. To drop the constant, he continued, would be "knocking the bottom out of space." Whereas the theoretical value of the cosmological constant is today one of physics' deep and still unsolved problems, to

Eddington it was not. In 1931 he provided an answer in terms of other constants of nature:

$$\Lambda = \left(\frac{2GM}{\pi}\right)^2 \left(\frac{mc}{e^2}\right)^4 = 9.8 \times 10^{-55} \text{ cm}^{-2}$$

While this was believed to be of roughly the right order, unfortunately (or fortunately for Eddington) there were no astronomical determinations of $\Lambda$ with which the theoretical value could be compared. Eddington used the value of $\Lambda$ to calculate from first principles the Hubble recession constant, for which he obtained $H_0$ = 528 km s$^{-1}$ Mpc$^{-1}$. Since the figure agreed nicely with the generally accepted observational value, he took it as evidence of the soundness of his theoretical approach (Kragh 2011, p. 97).

In his 1929 paper Eddington derived that the fine-structure constant was related in a simple way to constants of a cosmological nature, such as given by the expression

$$\alpha = \frac{2\pi mc R_E}{h\sqrt{N^*}}$$

He considered the cosmical number $N^*$ – the number of electrons and protons in the closed universe – to be a most important constant of nature. According to conventional physics and cosmology there was nothing special about the number, which might well have been different, but Eddington not only insisted that it was a constant, he also claimed that it could not have been different from what it is. Moreover, he claimed that he was able to deduce $N^*$ rigorously from theory, just as he was able to deduce the other constants of nature. The result was

$$N^* = 2 \times 136 \times 2^{256} \cong 3.15 \times 10^{79}$$

Notice that Eddington gave the number precisely. He counted a positron as minus one electron, and a neutron as one proton and one electron. Assuming that the total number of electrons equals the number of protons, he further derived a relation between two of the very large dimensionless constants:

$$\frac{e^2}{GmM} = \frac{2}{\pi}\sqrt{N^*}$$

This was the relation that he had vaguely suggested as early as 1923. For the mass ratio *M*/*m* between the proton and the electron, Eddington argued that it could be found from the ratio of the two roots in the equation

$$10x^2 - 136\omega x + \omega^2 = 0$$

The quantity $\omega$ is what Eddington called a "standard mass," the mass of an unspecified neutral particle. In this way he derived the theoretical value *M*/*m* = 1847.6 or nearly the same as the experimental value (Durham 2006, pp. 211-218).

In the 1930s a few scientists speculated for the first time that some of the constants of nature might not be proper constants but instead quantities that vary slowly in time (Kragh 2011, pp. 167-192). Eddington considered such ideas to be pure nonsense. In 1938 Dirac proposed a cosmological theory based on the radical assumption that *G* decreased according to

$$\frac{1}{G}\frac{dG}{dt} = -3H_0$$

However, Eddington (1939a) quickly dismissed Dirac's theory as "unnecessarily complicated and fantastic." He was not kinder to contemporary speculations that the speed of light might be a varying quantity: "The speculation of various writers that the velocity of light has changed slowly in the long periods of cosmological time … is nonsensical because a change in the velocity of light is self-contradictory" (Eddington 1946, p. 8). Nearly sixty years later the so-called VSL (varying speed of light) theory developed by João Magueijo revived the discussion of whether Eddington's objection was reasonable or not (Ellis and Uzan 2006; Kragh 2006).

**5. Fundamental theory**

The research project that Eddington pursued with such fervour and persistence during the last 15 years of his life resulted in a long series of scientific papers and a couple of important monographs. Some of his books were highly technical while others were of a philosophical nature and mostly oriented toward a general readership. To Eddington, the latter were no less important than the first. He followed his research programme in splendid isolation, apparently uninterested in the work done by other physicists in the tradition he had initiated. The isolated and closed nature of Eddington's research is confirmed by bibliometric studies based on the list of publications given by his biographer, the Canadian astronomer Allie Vibert Douglas (1956). Among the references in Eddington's 14 research papers on his unified theory in the period 1929-1944, no less than 70 per cent are to his own works. By comparison, the average self-reference ratio in physics papers in the period was about 10 per cent (Kragh and Reeves 1991).

The first major fruit of Eddington's efforts appeared in 1936 in the form of *Relativity Theory of Protons and Electrons* (RTPE) a highly mathematical and personal exposition of his ongoing attempt to create a new basis for cosmology and physics. During the following years he prepared a systematic account of his theory and its mathematical foundation, but *Fundamental Theory* only appeared after his death. The title was not Eddington's, but chosen but the mathematician Edmund Whittaker who edited Eddington's manuscript and supervised it to publication.

Whittaker had closely followed Eddington's work which fascinated him more from a philosophical than a physical point of view. Like most scientists he remained unconvinced about the physical soundness of the grand project. In an extensive review of RTPE, Whittaker (1937) likened Eddington to a modern Descartes, suggesting that Eddington's theory did not describe nature any better than the vortex theory of the French rationalist philosopher. Nonetheless, he described Eddington as "a man of genius." Whittaker was not alone in comparing Eddington to Descartes. According to the philosopher Charlie Broad (1940, p. 312): "For Descartes the laws of motion were deducible from the perfection of God, whilst for Eddington they are deducible from the peculiarities of the human mind." Moreover, "For both philosophers the experiments are rather a concession to our muddle-headedness and lack of insight."

Eddington's ambitious project of reconstructing fundamental physics amounted to a theory of everything. The lofty goal was to deduce all laws and, ultimately, all phenomena of nature from epistemological considerations alone, thereby establishing physics on an a priori basis where empirical facts were in principle irrelevant. In RTPE (1936, pp. 3-5) he expressed his ambition as follows:

> It should be possible to judge whether the mathematical treatment and solutions are correct, without turning up the answer in the book of nature. My task is to show that our theoretical resources are sufficient and our methods powerful enough to calculate the constants exactly – so that the observational test will be the same kind of perfunctory verification that we apply sometimes to theorems in geometry. … I

> think it will be found that the theory is purely deductive, being based on epistemological principles and not on physical hypotheses.

At the end of the book (p. 327) he returned to the theme, now describing his aim in analogy with Laplace's omniscient intelligence or demon appearing in the *Exposition du Système du Monde* from 1796. However, there was the difference that Eddington's demon was essentially human in so far that it had a complete knowledge of *our* mental faculties. He wrote:

> An intelligence, unacquainted with our universe, but acquainted with the system of thought by which the human mind interprets to itself the content of its sensory experience, should be able to attain all the knowledge of physics that we have attained by experiment. He would not deduce the particular events or objects of our experience, but he would deduce the generalisations we have based on them. For example, he would infer the existence and properties of radium, but not the dimensions of the earth.

Likewise, the intelligence would deduce the exact value of the cosmical number (as Eddington had done) but not, presumably, the value of Avogadro's number.

Eddington's proud declaration of an aprioristic, non-empirical physics was a double-edged sword. On the one hand, it promised a final theory of fundamental physics in which the laws and constants could not conceivably be violated by experiment. On the other hand, the lack of ordinary empirical testability was also the Achilles-heel of the theory and a main reason why most physicists refused taking it seriously. Eddington was himself somewhat ambivalent with regard to testable predictions and did not always follow his rationalist rhetoric. He could not and did not afford the luxury of ignoring experiments altogether, but tended to accept them only when they agreed with his calculations. If this were not the case he consistently and often arrogantly explained away the disagreement by putting the blame on the measurements rather than the theory. Generally he was unwilling to let a conflict between a beautiful theory and empirical data ruin the theory. "We should not," Eddington (1935, p. 211) wrote, "put overmuch confidence in the observational results that are put forward until they have been confirmed by theory."

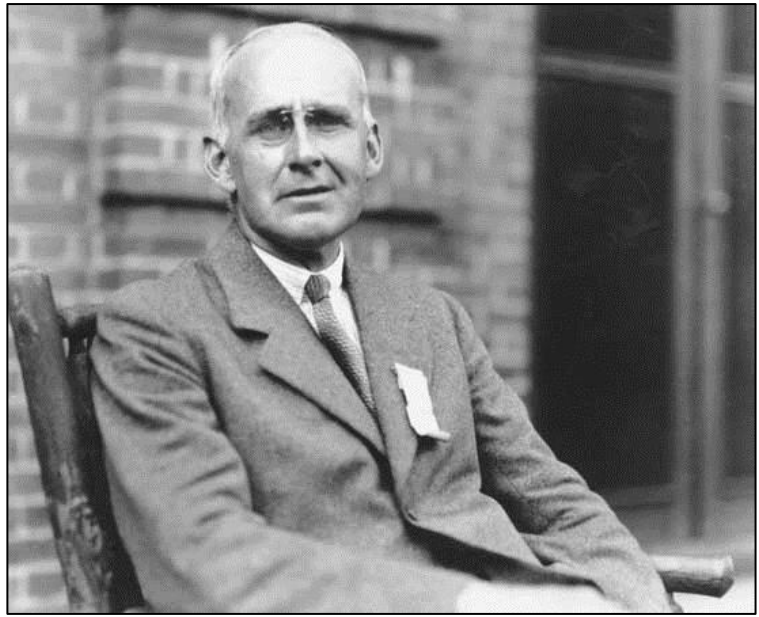

**Fig. 3.** Eddington in 1932. Credit: Encyclopædia Britannica.

## 6. Cosmo-physics

In order to understand Eddington's "flight of rationalist fancy" (Singh 1970, pp. 168-191) it is important to consider it in its proper historical context. If seen within the British tradition of so-called cosmo-physics his ambitious research project was not quite as extreme as one would otherwise judge it. A flight of rationalist fancy it was, but in the 1930s there were other fancies of the same or nearly the same scale. To put it briefly, there existed in Britain in the 1930s a fairly strong intellectual and scientific tradition that in general can be characterised as anti-empirical and pro-rationalist, although in some cases the rationalism was blended with heavy doses of idealism. According to scientists associated with this attempt to rethink the foundation of physical science, physics was inextricably linked to cosmology. In their vision of a future fundamental physics, pure thought counted more heavily than experiment and observation. The leading cosmo-physicists of the interwar period, or as Herbert Dingle (1937) misleadingly called them, the "new Aristotelians," were Eddington and E. Arthur Milne, but also Dirac, James Jeans and several other scientists held views of a roughly similar kind (Kragh 1982).

Although the world system of Milne, a brilliant Oxford astrophysicist and cosmologist, was quite different from the one of Eddington, on the methodological level Milne's system shared the rationalism and deductivism that characterized Eddington's system. Among other things, the two natural philosophers had in common that their ideas about the universe – or about fundamental physics – gave high priority to mathematical reasoning and correspondingly low priority to empirical facts. Milne, much like Eddington, claimed that the laws of physics could ultimately be obtained from pure reasoning and processes of inference. His aim was to get rid of all contingencies by turning the laws of nature into statements no more arbitrary than mathematical theorems. As Milne (1948, pp. 10-12) put it: "Just as the mathematician never needs to ask whether a constructed geometry is true, so there is no need to ask whether our kinematical and dynamical theorems are true. It is sufficient that they are free from contradictions."

Despite the undeniable methodological affinity between the views of Milne and Eddington, the Cambridge professor insisted that his ideas were wholly different from those of his colleague in Oxford. Eddington (1939a) either ignored Milne's theory or he criticized it as contrived and even "perverted from the start."

Dirac's cosmological theory based on the $G(t)$ assumption was directly inspired by the ideas of Milne and Eddington. His more general view about fundamental physics included the claim of a pre-established harmony between mathematics and physics, or what he saw as an inherent mathematical quality in nature. By the late 1930s Dirac reached the conclusion that ultimately physics and pure mathematics would merge into one single branch of sublime knowledge. In his James Scott Lecture delivered in Edinburgh in early 1939, he suggested that in the physics of the future there would be no contingent quantities at all. Even the number and initial conditions of elementary particles, and also the fundamental constants of nature, must be subjects to calculation. Dirac (1939, p. 129) proposed yet another version of Laplace's intelligence:

> It would mean the existence of a scheme in which the whole of the description of the universe has its mathematical counterpart, and we must assume that a person with a complete knowledge of mathematics could deduce, not only astronomical data, but also all the historical events that take place in the world, even the most trivial ones. … The scheme could not be subject to the principle of simplicity since it would have to be extremely complicated, but it may well be subject to the principle of mathematical beauty.

Note that Dirac's version included even "the most trivial" events in the world. This was not Eddington's view, for he believed that contingent facts – those "which distinguish the actual universe from all other possible universes obeying the same laws" – were "born continually as the universe follows its unpredictable course" (Eddington 1939b, p. 64). Another major difference between the two natural philosophers was Dirac's belief that the laws of physics, contrary to the rules of mathematics, are chosen by nature herself. This evidently contradicted Eddington's basic claim that physical knowledge is wholly founded on epistemological considerations.

## 7. Nature as a product of the mind

Although Eddington's project had elements in common with the ideas of Milne and other cosmo-physicists of the period, it was unique in the way he interpreted it philosophically. As mentioned in Section 2, Eddington was convinced that the laws of nature were subjective rather than objective. The laws, he maintained, were not summary expressions of regularities in an external world, but essentially the constructions of the physicists. This also applied to the fundamental constants of nature. Eddington (1939b, p. 57) characterized his main exposition of philosophy of physics, *The Philosophy of Physical Science*, as "a philosophy of subjective natural law." Referring to the cosmical number $N^*$, elsewhere in the book (p. 60) he explained that "the influence of the sensory equipment with which we observe, and the intellectual equipment with which we formulate the results of observation as knowledge, is so far-reaching that by itself it decides the number of particles into which the matter of the universe appears to be divided."

In agreement with his religious belief as a Quaker, Eddington deeply believed in an open or spiritual world that was separate from the one we have empirical access to (Stanley 2007). He often pointed out that physics is restricted to a small part of what we experience in a wider sense, as it applies only to what can be expressed quantitatively or metrically. "Within the whole domain of experience [only] a selected portion is capable of that exact representation which is requisite for development by the scientific method," he wrote in *The Nature of the Physical World*

(p. 275). Far from wanting physics to expand its power to the spiritual or non-metrical world, as others wanted it, Eddington found it preposterous to believe that this world could be ruled by laws like those known from physics or astronomy (Douglas 1956, p. 131). Given that his theory was limited to the metrical world, it was not really a theory of *everything*.

A key element in Eddington's epistemology was what he referred to as "selective subjectivism." With this term he meant that it is the mind which determines the nature and extent of what we think of as the external world. We force the phenomena into forms that reflect the observer's intellectual equipment and the instrument he uses, much like the bandit Procrustes from Greek mythology. The physicist, Eddington (1936, p. 328) wrote, "might be likened to a scientific Procrustes, whose anthropological studies of the stature of travellers reveal the dimensions of the bed in which he has compelled them to sleep." As a result of the selective subjectivism, "what we comprehend about the universe is precisely that which we put into the universe to make it comprehensible."

Eddington's anthropomorphic and constructivist view of laws of nature was related to the conventionalist view of scientists such as Karl Pearson and Henri Poincaré, only did it go much farther. Eddington (1935, p. 1) recognized the similarity to the view of the great French mathematician, from whose book *The Value of Science* he approvingly quoted: "Does the harmony which human intelligence thinks it discovers in Nature exist apart from such intelligence? Assuredly no. A reality completely independent of the spirit that conceives it, sees or feels it, is an impossibility."

The basic idea of the human mind as an active part in the acquisition of knowledge in the physical sciences, or even as the generator of the fabric of the cosmos, was not a result of Eddington's fundamental theory developed in the 1930s. More than a decade earlier, in the semi-popular *Space, Time and Gravitation* (p. 200), he wrote:

> We have found that where science has progressed the farthest, the mind has bur regained from nature that which the mind has put into nature. We have found a strange foot-print on the shores of the unknown. We have devised profound theories, one after another, to account for its origin. At last, we have succeeded reconstructing the creature that made the foot-print. And Lo! It is our own.

## 8. Quantum objections

Eddington's numerological and philosophical approach to fundamental physics attracted much attention among British scientists, philosophers and social critics in particular. The general attitude was critical and sometimes dismissive, as illustrated by the philosopher Susan Stebbing (1937), who in a detailed review took Eddington to task for what she considered his naïve philosophical views. He was, she said, a great scientist but an incompetent philosopher. According to the Marxist author Christopher Caudwell (1939, p. 121), Eddington was a scholastic who wanted to "extract truth by mathematical manipulation at the expense of experiment."

Leading theoretical physicists preferred to ignore the British astronomer-philosopher's excursion into unified physics rather than arguing with him. Many may have shared the view of Wolfgang Pauli, who in a letter of 1929 described Eddington's ideas as "complete nonsense" and "romantic poetry, not physics" (Kragh 2011, p. 109). Pauli referred specifically to Eddington's identification of the fine-structure constant $\alpha$ with the number 1/136. A main reason for the generally unsympathetic response to Eddington's theory in the physics community was his unorthodox use and understanding of quantum mechanics. I shall limit myself to some facets of this issue.

Eddington's critique of the standards employed in quantum mechanics generally fell on deaf ears among experts in the field. One of the few exceptions was a paper of 1942 in which Dirac, together with Rudolf Peierls and Maurice Pryce, politely but seriously criticized Eddington's "confused" use of relativistic quantum mechanics. As the three physicists pointed out: "Eddington's system of mechanics is in many important respects completely different from quantum mechanics … [and] he occasionally makes use of concepts which have no place there" (Dirac, Peierls and Pryce 1942, p. 193). The sharp difference between Eddington's quantum-cosmological theory and established quantum mechanics had earlier been highlighted at a conference on "New Theories in Physics" held in Warsaw and Cracow in June 1938. On this occasion Eddington (1939c) gave a lecture in front of some of the peers of orthodox quantum mechanics, including Niels Bohr, Léon Rosenfeld, Louis de Broglie, Oskar Klein, Hendrik Kramers, John von Neuman, George Gamow and Eugene Wigner (Fig. 4).

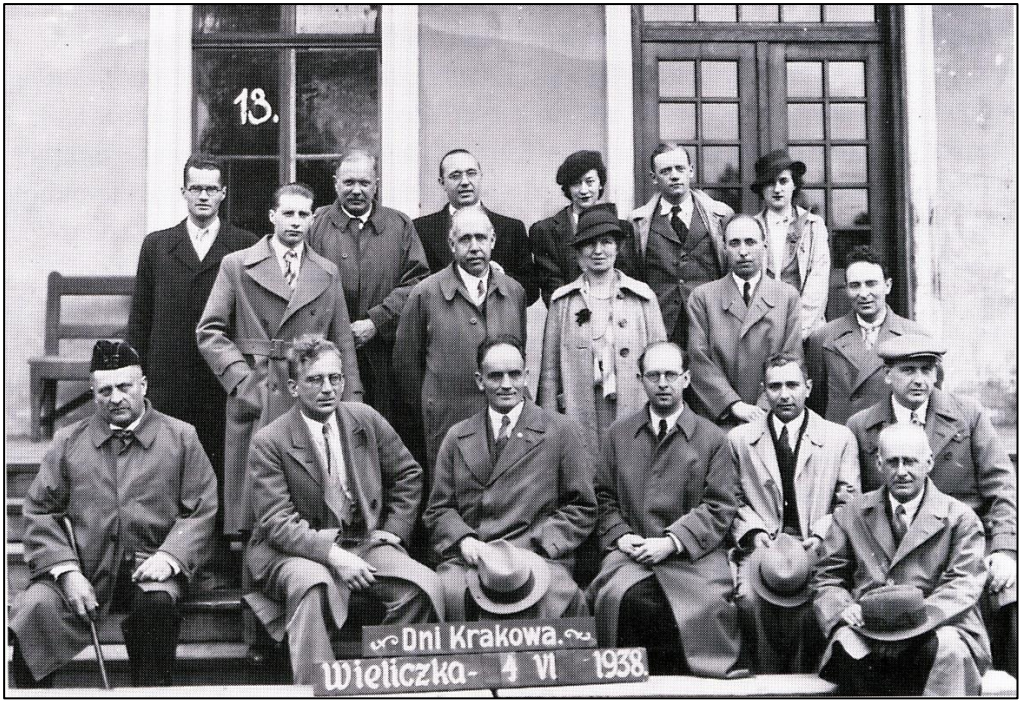

**Fig. 4.** Attendants at the Cracow session in 1938, with Eddington in the lower right corner. Reproduced from *Niels Bohr Collected Works*, vol. 7 (Amsterdam: Elsevier, 1996), p. 261.

None of the distinguished quantum physicists could recognize in Eddington's presentation what they knew as quantum mechanics. Kramers commented: "When listening to Prof. Eddington's interesting paper, I had the impression that it concerned another quantum theory, in which we do not find the formulae ordinarily used, but where we find many things in contradiction with the ordinary theory." In the proceedings of the Polish conference one gets a clear impression of how Eddington on the one hand, and Bohr and his allies on the other, failed to communicate. It was one paradigm challenging another, apparently incommensurable paradigm. The attempt to create a dialogue between Bohr and Eddington led to nothing. According to the proceedings, Bohr "thought that the whole manner of approaching the problem which Professor Eddington had taken was very different from the quantum point of view." And Eddington, on his side, stated that "he could not understand the attitude of Prof. Bohr." He somewhat lamely, and somewhat pretentiously, responded that he just tried to do for quantum mechanics what Einstein had done for classical, non-quantum mechanics.

Eddington realized that he was scientifically isolated, yet he felt that the lack of appreciation of his ideas was undeserved and would change in the future. Near

418the end of his life he confided in a letter to Dingle (1945, p. 247) that he was perplexed that physicists almost universally found his theory to be obscure. He defended himself: "I cannot seriously believe that I ever attain the obscurity that Dirac does. But in the case of Einstein and Dirac people have thought it worth while to penetrate the obscurity. I believe they will understand me all right when they realize that they have got to do so."

Although the large majority of physicists dismissed Eddington's theory there was one notable exception, namely Erwin Schrödinger. In papers from the late 1930s the father of wave mechanics enthusiastically supported Eddington's quantum-cosmological theory (Kragh 1982; Rüger 1988). Yet, his enthusiasm cooled as it dawned upon him that the theory could not be expressed in a language accessible to the physicists. In an essay originally written in 1940 but only published much later, Schrödinger (1953, p. 73) admitted that an important part of Eddington's theory "is beyond my understanding."

Still today this is the general verdict of Eddington's grand attempt to establish fundamental physics on an entirely new basis.

**References**

Barrow, John D. (2002). *The Constants of Nature: From Alpha to Omega*. London: Jonathan Cape.
Broad, Charlie (1940). "Discussion of Sir Arthur Edington's 'The Philosophy of Physical Science'." *Philosophy* **15**, 301-312.
Caudwell, Christopher (1939). *The Crisis in Physics*. London: Bodley Head.
Davis, Bergen (1904). "A suggestive relation between the gravitational constant and the constants of the ether." *Science* **19**, 928-929.
Dingle, Herbert (1937). "Modern Aristotelianism." *Nature* **139**, 784-786.
Dingle, Herbert (1945). "Sir Arthur Eddington, O.M., F.R.S." *Proc. Phys. Soc.* **57**, 244-249.
Dirac, Paul A. M. (1939). "The relation between mathematics and physics." *Proc. Roy. Soc. (Edinburgh)* **59**, 122-129.
Dirac, Paul A. M., Rudolf Peierls, and Maurice Pryce (1942). "On Lorentz invariance in the quantum theory." *Proc. Cambr. Phil. Soc.* **38**, 193-200.
Durham, Ian (2006). "Sir Arthur Eddington and the foundation of modern physics." Arxiv:quant-ph/0603146.
Douglas, Allie V. (1956). *The Life of Arthur Stanley Eddington*. London: Thomas Nelson & Sons.
Eddington, Arthur S. (1920a). *Space, Time and Gravitation: An Outline of the General Relativity Theory*. Cambridge: Cambridge University Press.